# LCDctCNN: Lung Cancer Diagnosis of CT scan Images Using CNN Based Model

Muntasir Mamun
*Department of Computer Science*
*University of South Dakota*
South Dakota, USA
muntasir.mamun@coyotes.usd.edu

Md Ishtyaq Mahmud
*College of Science & Engineering*
*Central Michigan University*
Mount Pleasant, MI 48858, USA
mahmu4m@cmich.edu

Mahabuba Meherin
*Department of Computer Science & Engineering*
*American International University-Bangladesh*
Dhaka, Bangladesh
meherin88@gmail.com

Ahmed Abdelgawad
*College of Science & Engineering*
*Central Michigan University*
Mount Pleasant, MI 48858, USA
abdel1a@cmich.edu

*Abstract*— The most deadly and life-threatening disease in the world is lung cancer. Though early diagnosis and accurate treatment are necessary for lowering the lung cancer mortality rate. A computerized tomography (CT) scan-based image is one of the most effective imaging techniques for lung cancer detection using deep learning models. In this article, we proposed a deep learning model-based Convolutional Neural Network (CNN) framework for the early detection of lung cancer using CT scan images. We also have analyzed other models for instance Inception V3, Xception, and ResNet-50 models to compare with our proposed model. We compared our models with each other considering the metrics of accuracy, Area Under Curve (AUC), recall, and loss. After evaluating the model's performance, we observed that CNN outperformed other models and has been shown to be promising compared to traditional methods. It achieved an accuracy of 92%, AUC of 98.21%, recall of 91.72%, and loss of 0.328.

## I. Introduction

Lung cancer is one of the most deadly and devastating types of cancer in the world. It is challenging to detect cancer, and its symptoms only become noticeable in the final stages. Although this cancer's death rate could be decreased by early detection and appropriate treatment for patients. Lung cancer often starts in the lungs; however, it occasionally appears as early symptoms prior to spread [1]. In recent years, numerous techniques have been developed, and research is ongoing to effectively identify lung cancer. The greatest imaging method for early diagnosis of lung cancer will be CT scan images, although it can be challenging for medical professionals to interpret and detect cancer from CT scan images. [2].

Figure 1 depicts expected statistical information for a few cancer types in 2022. Making this figure we used the statistical data of the American Cancer Society (ACS) [3]. Based on ACS, the death rate of lung cancer is higher than any other cancer, which is around 0.13 million all over the world. Every year, there are a lot of new cases, with an estimated 0.237 million cases in 2022. The mortality rate is significant in the absence of appropriate treatment because this cancer is only discovered in its advanced stages and the ratio of new cases and the death rate is higher than any other cancer.

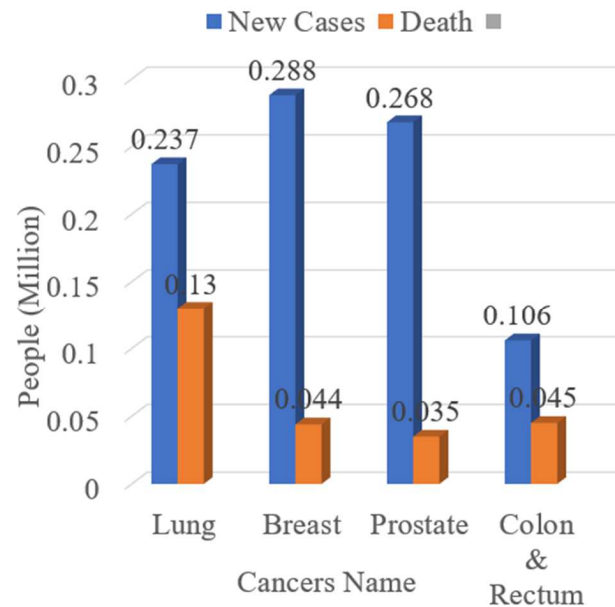

Figure 1: Cancer in 2020 (New cases against death)

Lung cancer cells can take the following forms: adenocarcinoma, large cell carcinoma, and squamous cell carcinoma. Adenocarcinoma is the most common cell of lung cancer which is found on the outer surface of the lung. Lung cancer with large-cell undifferentiated carcinoma has a rapid growth and dissemination rate and can occur at any place in the lung. Squamous cell carcinoma is cancer that relates to smoking and is located in the central region of the lung [4, 25].

To detect cancer, predict the outcome of cancer treatment, and increase patient survival after a diagnosis of cancer, a variety of methods are being explored. Techniques for screening, identifying, and classifying cancers have been utilized by medical professionals and researchers to make early cancer diagnoses. Nowadays, the machine learning model is widely used for detecting, analyzing, and classifying critical medical healthcare treatment [5, 6]. Convolutional Neural Network (CNN) based machine learning model can be the best for early detection, observing, and classifying lung cancer using CT scan images.

## II. Literature Review

Ausawalaithong et al. [7] utilized a convolutional neural network (CNN) to analyze a very big dataset of chest x-ray images to find anomalies. The authors evaluated the performance of the models using three retrained models and diverse datasets for accuracy, specificity, and sensitivity. Using the ChestX-ray14 dataset, Model A identified lung nodules. Model C identified lung cancer using both ChestX-ray14 and JSRT, and even though it had a lower standard deviation across all assessment parameters, it correctly identified the lung cancer's location. Model B displayed greater specificity but lower accuracy and sensitivity than Model C, and it did this using the dataset from the Japanese Society of Radiological Technology (JSRT). The authors suggested retraining the model numerous times for particular tasks. Since Model C correctly predicted the site of cancer while Retrained Model B produced unfavorable findings, it provides superior outcomes in virtually all metrics and can address the issue of a short dataset.

Bhandary et al. [8] utilized a modified version of AlexNet's (MAN) deep learning method to find lung anomalies including cancer and pneumonia. The authors used two types of datasets for accurate results. The datasets are Chest X-Ray and LIDC-IDR. The developed MAN-SVM technique was tested on the same dataset as the initial experimental investigation with AlexNet and provided the highest accuracy of 96.80% compared to all other techniques, 86.95% accuracy of less than 87% after modifications were implemented in the final stage of the DL structure. Again, a comparable DL architecture demonstrated a 97.27% accuracy.

Da Silva et al. [9] used CNN configuration generated by the PSO algorithm. To enable an accurate comparison between the particles, it was trained and validated on identical sets. The authors collected data from the LIDC-IDRI dataset. Five test subsets were used to obtain the results. Test-1 produced results of 96.54% accuracy, 87.79% sensitivity, 98.215% specificity, and 0.931 AUC with 17,870 samples used. Test-4, one of the five test subsets, produced the best results, scoring 97.62% accurate, 92.20% sensitive, 98.64% specific, and 0.955 AUC.

Naqi et al. [10] used Stacked Autoencoder + Softmax. The authors suggest classifying nodules combining data from 2D and 3D resources. For feature reduction and nodule categorization, deep learning is used. The LIDC-IDRI data set, which is accessible to the general public, is used for the experiment. The main evaluation criteria for this study are the performance aspects, which include sensitivity, specificity, accuracy, and a number of FPs/scans. The authors includes a total of 888 CT scans with 777 sizes ⩾ 3 mm nodules that have been identified by all four expert radiologists. The suggested method provided low false positive rates of 2.8/scan with 95.6% sensitivity, 96.9% accuracy, and 97.0% specificity, greatly improving the results.

Shaffie et al. [11] used Deep autoencoder. It introduces a newly developed automated noninvasive clinical diagnostic methodology for the early identification of lung cancer by identifying the benign or malignant nature of the observed lung nodule. The authors used the LIDC-IDRI data set and their system got promising results. The performance characteristics of this study were sensitivity, specificity, accuracy, and AUC. The suggested framework has the potential to aid in the early detection of lung cancer, with accuracy, specificity, sensitivity, and AUC values of 91.20%, 95.88%, 85.03%, and 95.73 obtained from a collection of 727 nodules collected from 467 individuals.

Kaur et al. [12] utilized CNN, the proposed CNN network consists of three sets of rectified linear unit (ReLU) layers, convolutional layers, CNN, and three sets of ReLU layers, followed by a fully connected layer. Each convolutional layer's representative features are retrieved via 64 filters. For training and validation, the Japanese Society of Radiological Technology (JSRT) dataset is utilized. The average accuracy, overlap, sensitivity, and specificity for the results were 98.05%, 96.25%, and 98.80%, respectively.

Xie et al. [13] utilized a deep neural network model called the multi-view knowledge-based collaborative (MV-KBC) for the categorization of benign and malignant lung nodules on chest CT. They evaluated their methodology against the five cutting-edge classification approaches using the reference LIDC-IDRI data set. According to their findings, the MV-KBC model classified lung nodules with an accuracy of 91.60% and an AUC of 95.70% and they suggested that their method might be applied in a routine clinical workflow.

Nibali et al. [14] applied the fundamental ideas of the ResNet to the categorization of lung nodules. The publicly accessible LIDC/IDRI dataset served as the foundation for the experimental dataset. 1010 patients' CT scans are stored in the LIDC, and four radiologists evaluated each patient's scan to generate four sets of arbitrary nodule readings. Two primary sets of findings are presented by the authors. First, the authors focused on the beneficial effects that curriculum learning, transfer learning, and deeper networks had on the system's classification accuracy. Second, using identical training and testing settings, authors objectively compared their system to versions of other models. The findings revealed that their approach outperformed all others in terms of sensitivity (91.07%), specificity (88.64%), precision (89.35%), AUROC (0.9459), and accuracy (89.90%).

Zhang et al. [15] used a deep belief network (DBN). The authors also used the publicly available dataset LIDC-IDRI. The algorithm's accuracy, sensitivity, and specificity were all above 90% when used to detect big nodules >30 mm. The articles had a sensitivity range of 84.2%, and 87.1%, and an accuracy range of 89.0%–89.5%.

Causey et al. [16] presented NoduleX, a practical method for determining the malignancy of lung nodules from patients' CT images that uses deep learning convolutional neural networks (CNN). For training and validation, the scientists examined the nodules from the LIDC/IDRI cohort. They found that NoduleX can achieve 0.99 AUC on the independent validation test with an accuracy of 94.6%, sensitivity of 94.8%, and specificity of 94.3%.

TABLE I. Lung cancer detection model's performance

| Authors (year) | Applied Models | Dataset Collection (Image samples) | Measures (Proposed model) |
|---|---|---|---|
| Ausawalaithong et al. (2018) [7] | CNN | ChestX-ray14 (30,805) | Accuracy: 84.02% Specificity: 85.34% Sensitivity: 82.71% |
| | | JSRT (247) | Accuracy: 65.51 ± 7.67% Specificity: 80.95 ± 20.59% Sensitivity: 45.67 ± 21.36% |
| | | ChestX-ray14 and JSRT | Accuracy: 74.43 ± 6.01% Specificity: 74.96 ± 9.85% Sensitivity: 74.68 ± 15.33% |
| Bhandary et al. (2019) [8] | Modified AlexNet (MAN) | Chest X-Ray | Accuracy: 96.80% Specificity: 96.63% Sensitivity: 96.97% F1 score: 96.78% |
| | | LIDC-IDR | Accuracy: 97.27% Specificity: 95.63% Sensitivity: 98.09% F1 score: 97.95% |
| Da Silva et al. (2017) [9] | CNN | LIDC-IDRI (18,408) | Accuracy: 97.62% Specificity: 98.64% Sensitivity: 92.20% AUC: 95.5% |
| Naqi et al. (2018) [10] | Stacked Autoencoder + Softmax | LIDC-IDRI (888) | Accuracy: 96.9% Sensitivity: 95.6% Specificity: 97.0% |
| Shaffie et al. (2018) [11] | Deep autoencoder | LIDC-IDRI (727) | Accuracy: 91.20% Specificity: 95.88% Sensitivity: 85.03% |
| Kaur et al. (2017) [12] | CNN | JSRT (247) | Accuracy: 98.05% Sensitivity: 96.25% Specificity: 98.80% Overlap: 93.4% |
| Xie et al. (2019) [13] | MV-KBC | LIDC-IDRI | Accuracy: 91.6% Sensitivity: 86.5% Specificity: 94.0% AUC: 95.73% |
| Nibali et al. (2017) [14] | ResNet | LIDC-IDRI | Accuracy: 89.9% Sensitivity: 91.1% Specificity: 88.6% |
| Zhang et al. (2017) [15] | DBN | LIDC-IDRI | Accuracy: 95.0% Sensitivity: 93.5% Specificity: 90.2% AUC: 93% |
| Causey et al. (2018) [16] | CNN | LIDC-IDRI (1018) | Accuracy: 94.6 % Sensitivity: 94.8 % Specificity: 94.3 % AUC: 0.984 % |
| **Our work (2022)** | CNN, ResNet-50, Inception V3, Xception | Kaggle- CT scans (967) | Accuracy: 92% Recall: 91.72% AUC: 98.21% Loss: 0.328 |

### III. METHODOLOGY

The methodology begins with an image dataset obtained from a publicly available source. The image dataset is then pre-processed. The proposed CNN model, as well as other deep learning models such as ResNet-50, Inception V3, and Xception, are then trained, tested, and validated on the computerized tomography (CT) scan dataset using the standard hold-out-validation method. The results are computed and analyzed to determine the best deep learning-based model for detecting lung cancers such as adenocarcinoma, large cell carcinoma, and squamous cell carcinoma, as well as normal (not lung cancer). CNN is a custom-trained model, whereas ResNet-50, Inception V3, and Xception are pre-trained transfer learning models [17,18,21,23,24]. As a result, Figure 3 depicts the proposed custom CNN architecture, while Figure 2 depicts an overview of the proposed strategy.

#### A. Dataset collection:

Here, the lung cancer Dataset (CT scan Images) has been collected from the publicly available "Kaggle" online source [4]. According to the dataset source, the images were hand collected from various websites, with each and every label verified. Images are not in DCM format, the images are in JPG or PNG to fit the model. The data consists of 967 CT scan images. The dataset has four types of classes: adenocarcinoma, large cell carcinoma, squamous cell carcinoma, and normal (not lung cancer) for diagnosing lung cancer.

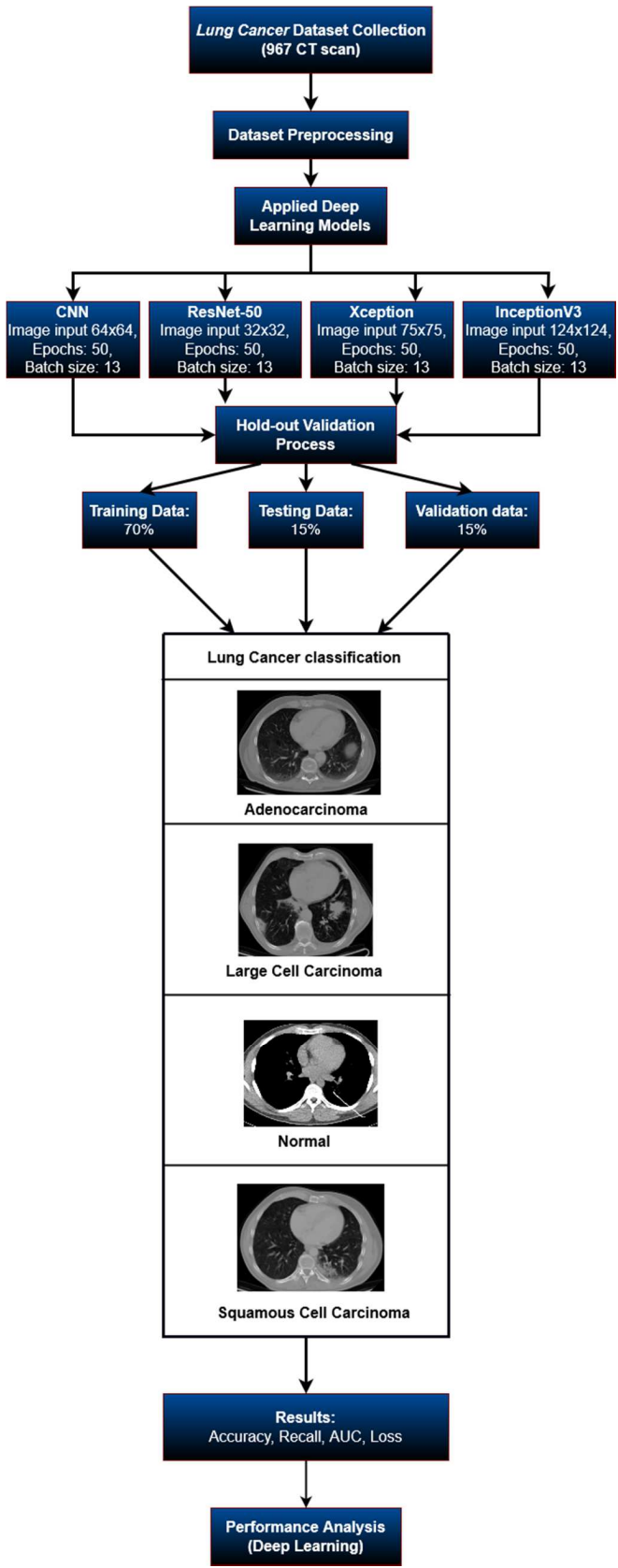

Figure 2: Overview of the study.

### B. Dataset pre-processing:

The images were pre-processed using feature extraction, which included reading the images, resizing them, removing noises (de-noise), image segmentation, and morphology (smoothing edges). This processing system is essential for analyzing deep learning models for image classification or detection.

### C. Validation process:

For large image datasets, it is critical to choose the best validation procedure. We used a hold-out validation process, keeping 70% of the data for training, 15% for testing, and 15% for validating. The hold-out validation technique is the most commonly used method and produces effective results [19]. For all the deep learning models, we selected the epochs value of 50 and batch size value of 13. We also used a random seed value of 1000 while implementing all the models, so that we can get the re-producible implemented results, or else the results would change in every iteration.

### D. Proposed CNN architecture:

The 64x64 input image was first sent to a first convolution layer in the proposed CNN, which has a value of 16 filters and 62x62 feature maps to look for the most fundamental features. The convolutional layer was the main building block of CNN. The output of the convolutional layer was then passed on to a max pooling layer with feature maps of 31x31 in order to reduce the size of the spatial data for the subsequent layer by half. Max pooling selects the maximum elements or pixels from the area of the feature map covered by the filter. Then, for additional processing, this output was sent to a second convolution layer with a value of 32 filters and 29x29 feature maps. The output of this layer was then passed on to a max pooling layer with 14x14 feature maps in order to decrease the amount of spatial data for the subsequent layer in half. Another set of convolution and pooling layers was added in the third step. In this case, the pooling layer consisted of 5x5 feature maps and the convolution layer was consisting of 64 filters with 10x10 feature maps. Then, the final output was flattened and moved to the 260-dimensional fully connected dense layer. After that, it is routed to the activation function layer which was softmax. Softmax activation function is generally used for multiple classifications. Except for the final layer, all layers used a ReLU activation function with no dropout. Figure 3 depicts the proposed CNN architecture's above-mentioned layout. With a learning rate of 0.01, 50 epochs, and 13 batch sizes, the model was trained, validated, and tested. The model was compiled using the Adam optimizer. Using the Keras Python library, a categorical cross-entropy based loss function and other metrics such as accuracy, recall, and AUC were achieved.

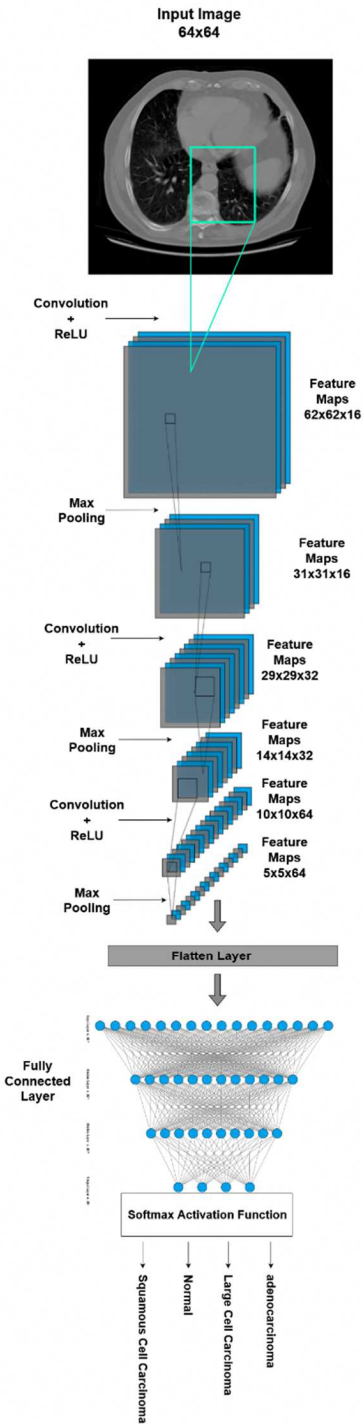

Figure 3: CNN architecture

## IV. RESULTS AND DISCUSSION

The findings of four types of deep learning models - i.e. CNN, ResNet-50, Inception V3, and Xception classification algorithms on the Lung cancer CT scan image dataset have been computed in table II, table III, table IV and comparisons have been provided in Figure 4. In table II, table III, and table IV, we presented the results for training, validation, and testing output for the performance observation of the models respectively. After analyzing the methods of CNN, ResNet-50, Inception V3, and Xception, it is observed that CNN outperforms other deep learning models based on the findings in table II, table III, and table IV. CNN considered the proposed model for lung cancer detection by CT scan images. The CNN achieved a testing accuracy of 92%, a testing AUC of 98.21%, testing recall of 91.72%, and a testing loss of 0.328.

Table II. Training results for different deep learning models for detecting Lung cancer.

| Models | Training Accuracy | Training AUC | Training Recall | Training Loss |
|---|---|---|---|---|
| CNN | 99.80% | 100% | 99.70% | 0.002 |
| ResNet-50 | 99.56% | 99.99% | 99.60% | 0.045 |
| Inception V3 | 94.25% | 96.40% | 94.23% | 1.960 |
| Xception | 94.10% | 96.70% | 94.08% | 1.450 |

Table III. Validation results for different deep learning models for detecting Lung cancer.

| Models | Val. Accuracy | Val. AUC | Val. Recall | Val. Loss |
|---|---|---|---|---|
| CNN | 91.10% | 97.80% | 91.03% | 0.352 |
| ResNet-50 | 84.20% | 94.90% | 83.50% | 0.598 |
| Inception V3 | 82.07% | 88.50% | 82.10% | 15.70 |
| Xception | 82.10% | 90.00% | 82.06% | 8.270 |

Table IV. Testing results for different deep learning models for detecting Lung cancer.

| Models | Testing Accuracy | Testing AUC | Testing Recall | Testing Loss |
|---|---|---|---|---|
| CNN | 92.00% | 98.21% | 91.72% | 0.328 |
| ResNet-50 | 84.13% | 94.85% | 83.45% | 0.598 |
| Inception V3 | 82.07% | 88.50% | 82.06% | 15.70 |
| Xception | 82.10% | 90.00% | 82.07% | 8.270 |

Figure 4 illustrates the comparison of the model's accuracy, AUC, and loss. Accuracy, AUC, and loss have all been taken into consideration when evaluating the models' performance. Figure 4 shows that compared to other models, CNN had the highest testing accuracy which is 92%. ResNet-50, Inception V3, and Xception achieved testing accuracy of 84.13%, 82.07%, and 82.10% respectively.

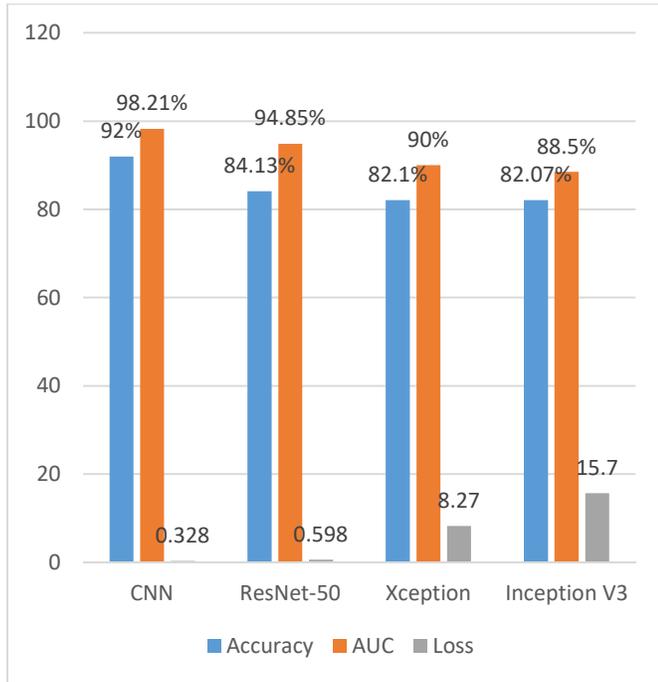

Figure 4: Deep Learning model's performance analysis in terms of accuracy, AUC, and loss.

AUC is also a crucial metric for evaluating the model's performance. AUC determines the model's performance and assesses a model's ability to differentiate between classes. The AUC measures how well the model differentiates between positive and negative classes. The higher the AUC value, the better the model's performance. The value range is 0 to 1, with 0 representing an incorrect test and 1 representing an accurate test. In general, an AUC of 0.5 indicates no discrimination (i.e., the ability to classify lung cancer), 0.7 to 0.8 is considered acceptable, 0.8 to 0.9 is considered great performance, and greater than 0.9 is considered outstanding performance [20]. Based on figure 4, CNN not only achieved the highest testing accuracy only but also achieved the highest test AUC score which is 98.21%. In addition, ResNet-50, Xception, and Inception V3 achieved testing AUC scores of 94.85%, 90%, and 88.5% respectively. However, the loss is another important metric for considering the model's performance. Loss is a number that indicates how inaccurate the model's prediction is at each epoch. If the loss is zero the model's prediction is perfect; otherwise, the greater the loss worse the model's performance. To calculate the loss in the detecting process, we used the categorical cross-entropy loss function. Categorical cross-entropy is a loss function that is mostly used in multi-class classification tasks [21]. Here, CNN achieved the lowest loss value of 0.328 and ResNet-50 achieved the loss value of 0.598. On the contrary, Xception and Inception V3 models achieved very high loss values which are 8.27 and 15.7 respectively. . From the above analysis, the custom CNN model outperforms other deep learning models in detecting different types of lung cancer using the CT scan image dataset and considered the proposed model. We provided the validation accuracy, validation AUC and loss function curve with respect to training accuracy, training AUC, and training loss in every epoch for CNN in Figures 5, 6, and 7, respectively.

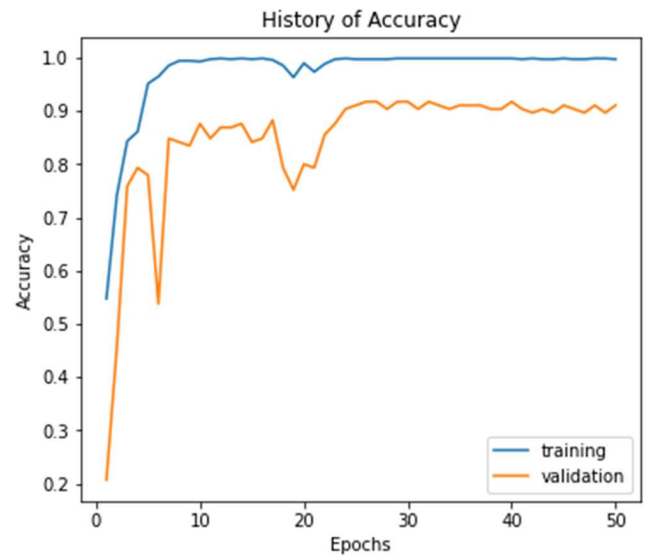

Figure 5: Accuracy curve for CNN

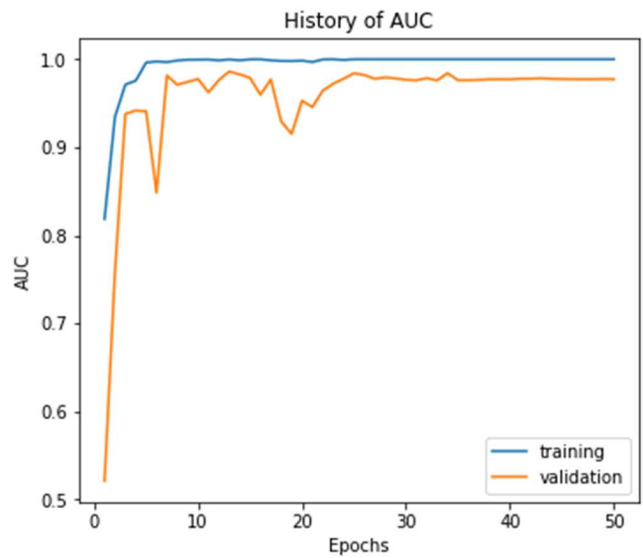

Figure 6: AUC curve for CNN

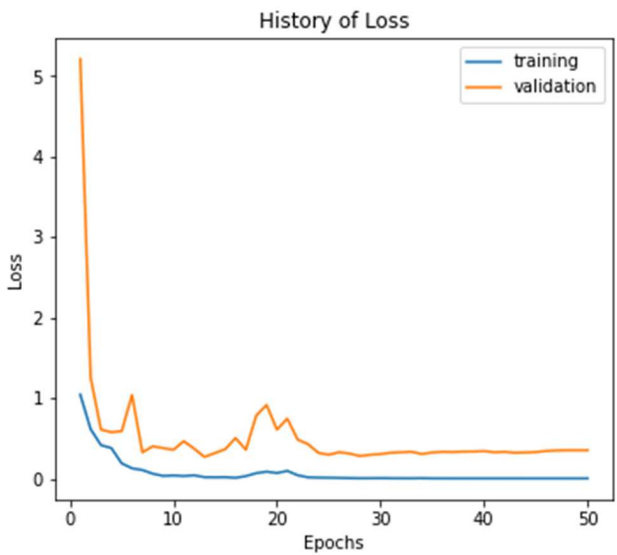

Figure 7: Loss function curve for CNN

## V. Conclusion and Future work

The death rate for lung cancer is shown in Figure 1 and indicates that it is one of the most prevalent and leading cancers worldwide. Although it cannot be prevented, a quick diagnosis can help the patient live longer than expected. In North America and other industrialized nations, lung cancer is the primary reason for cancer-related mortality. Lung cancer is at the top of the priority list because it frequently isn't discovered until the disease is well along. Therefore, despite substantial advancement over the past years, early diagnosis is still not reliable. In our research, we proposed CNN based deep learning model for the early detection of lung cancer using CT scan images. If we detect early-stage in cancer, it might be possible to cure cancer with proper treatment and care. Along with the CNN model, we also analyzed others model such as ResNet50, Inception V3, and Xception. We found that CNN outperformed other models with an accuracy of 92%, AUC of 98.21%, recall of 91.72%, and loss of 0.328 after analyzing our deep learning model using CT scan images. In order to diagnose lung cancer early, we may in the future take into account more datasets and various machine learning and deep learning models for better performance.